\documentclass[a4paper]{jpconf}
\usepackage{graphicx}
\begin{document}
\title{The Superconducting Gap Behavior in the Antiferromagnetic Nickel-Borocarbide
Compounds $R$Ni$_2$B$_2$C ($R$=Dy, Ho, Er, Tm) Studied by
Point-Contacts Spectroscopy}

\author{Yu. G. Naidyuk$^1$,  G. Behr$^2$, N. L. Bobrov$^1$, V. N. Chernobay$^1$, S.-L.
Drechsler$^2$, G. Fuchs$^2$, O. E. Kvitnitskaya$^{1,2}$ D. G.
Naugle$^3$, K.~D.~D.~Rathnayaka$^3$ and I. K. Yanson$^1$}

\address{$^1$ B. Verkin Institute for Low Temperature Physics and Engineering,
National Academy of Sciences of Ukraine, 47 Lenin ave., 61103,
Kharkiv, Ukraine}

\address{$^2$ Leibniz-Institut f\"ur Festk\"orper- und Werkstoffforschung Dresden,
POB 270116, D-01171 Dresden, Germany}

\address{$^3$ Department of Physics, Texas A\&M University, College Station TX 77843-4242, USA }

\ead{naidyuk@ilt.kharkov.ua}

\begin{abstract}
An general survey of the superconducting (SC) gap study in the
title compounds by point-contact (PC) spectroscopy is presented.
The SC gap was determined from $dV/dI$ of PCs employing the
well-known theory of conductivity for normal metal-superconductor
PCs accounting Andreev reflection.  The theory was modified by
including pair-breaking effects considering the presence of
magnetic rare-earth ions. A possible multiband structure of these
compounds was also taken into account. The PC study of the gap in
the Er--compound ($T_{\mbox{\tiny N}}\simeq$ 6\,K $< T_{\rm
c}\simeq$ 11\,K) gives evidence for the presence of two SC gaps.
Additionally, a distinct decrease of both gaps is revealed for $R$
= Er in the antiferromagnetic (AF) state. For $R$ = Tm
($T_{\mbox{\tiny N}}\simeq$1.5\,K $< T_{\rm c}\simeq$10.5\,K) a
decrease of the SC gap is observed below 4--5\,K, while for $R$ =
Dy ($T_{\mbox{\tiny N}}\simeq$10.5\,K $>T_{\rm c}\simeq$ 6.5\,K)
the SC gap has a BCS-like dependence in the AF state. The SC gap
for $R$ = Ho ($T_{\mbox{\tiny N}}\simeq$5.2\,K $< T_{\rm c}\simeq$
8.5\,K) exhibits below $T^*\simeq$5.6\,K a single-band BCS-like
dependence vanishing above $T^*$, where a specific magnetic order
occurs. The difference in the SC gap behavior in the title
compounds is attributed to different AF ordering.
\end{abstract}

\section{Introduction}
The $R$Ni$_2$B$_2$C ($R$=Dy, Ho, Er, Tm) family of ternary
superconductors attracts attention \cite{Muller08} because
superconductivity and long-range antiferromagnetic (AF) order
compete in these materials and their superconducting (SC)
properties exhibit often unconventional behavior. It turned out
also that a multiband approach is required to describe properly SC
properties of $R$=Y and Lu nickel borocarbides as it was shown in
\cite{Shulga}. Additionally, the AF order has its own specific
features in each of the title compounds, e.\,g., along with
different N\`eel temperature $T_{\mbox{\tiny N}}$ the AF order can
be in- or commensurate \cite{Muller08}. The manifestation of the
mentioned extraordinary behavior of $R$Ni$_2$B$_2$C is intimately
dependent on the chemical composition and crystal perfectness,
therefore continuous progress in synthesis of high quality single
crystal samples leads to deeper understanding of their fundamental
physics. In this paper we overview the efforts in point-contact
(PC) studies of the SC gap in the title compounds based on very
recent research (see \cite{Bobrov,Naid07,Naid07a,Bobrov08} and
Refs. therein). The PC method \cite{Naid} allows to study the
directional, temperature and magnetic field dependence of the SC
gap which can provide insight into the SC ground state of
$R$Ni$_2$B$_2$C and its interplay with magnetic order.

\section{Experimental}
Here we will focus on results measured on the best single crystals
reported so far. HoNi$_2$B$_2$C ($T_{\mbox{\tiny N}} \approx $
5.2\,K $< T_{\rm c}\approx $ 8.5\,K) and TmNi$_2$B$_2$C
($T_{\mbox{\tiny N}} \approx $ 1.5\,K $< T_{\rm c} \approx $
10.5\,K) crystals were grown by a floating zone technique with
optical heating at IFW Dresden, while ErNi$_2$B$_2$C
($T_{\mbox{\tiny N}}\approx $ 6\,K $< T_{\rm c} \approx $ 11\,K)
and DyNi$_2$B$_2$C ($T_{\mbox{\tiny N}}\approx $ 10.5\,K $
> T_{\rm c} \approx $ 6.5\,K) were grown at Ames Laboratory
(USA) by a Ni$_2$B high-temperature flux growth method. PCs were
established at the liquid helium temperature by standard
"needle-anvil" methods \cite{Naid}, using as a "needle" a
sharpened piece of Cu or Ag. The SC gap was evaluated from the
measured $dV/dI(V)$ dependences of PCs applying both the standard
BTK theory and an akin theory (in the case of Er), which
considered the pair-breaking effect of magnetic impurities
\cite{Belob} \footnote{The theory \cite{Belob} was used to fit
$dV/dI(V)$ for $R$=Er PCs. In this case instead of the SC gap the
SC order parameter (OP) $\Delta $ is evaluated while the SC gap
$\Delta _0=\Delta $(1-$\gamma ^{2/3})^{3/2}$, here $\gamma $ is
pair-breaking parameter. Forasmuch as $\gamma $ variates at our OP
evaluation below 0.1, the gap has similar behavior like OP, shown
in Fig.\,3 for $R$=Er. Therefore, for uniformity we used
definition "one-gap" or "two-gap" fit, although in the case of
$R$=Er a notation "two-OP" fit is more adequate.}. This is
reasonably for the mentioned borocarbides because of the presence
of magnetic moments on the rare-earth ions.

\vspace{2cm}
\begin{figure}[h]
\begin{minipage}{18pc}
\includegraphics[width=18pc]{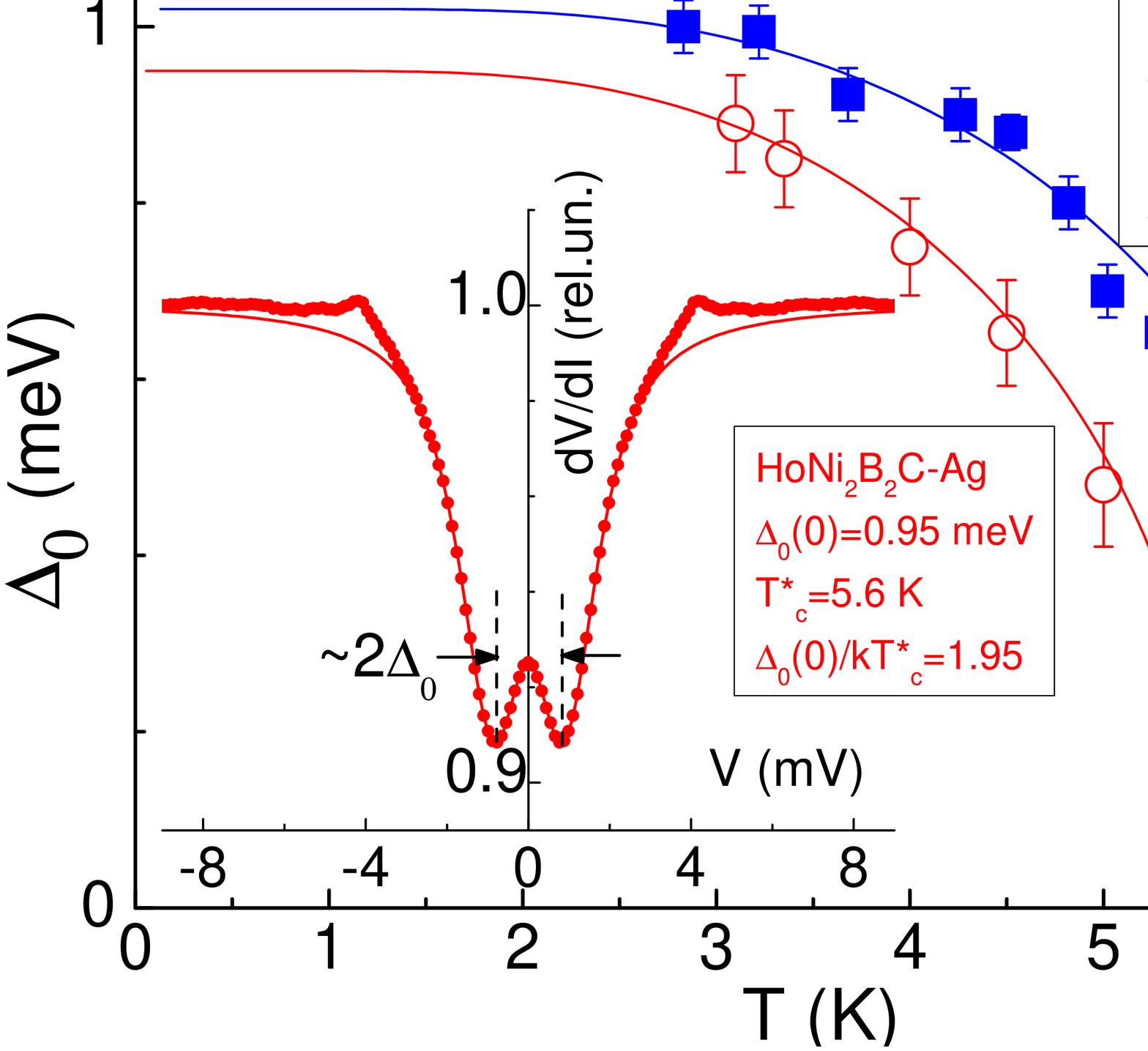}
\vspace{-3cm} \caption{\label{f1}Gap behavior in HoNi$_2$B$_2$C
(circles) \cite{Naid07} and DyNi$_2$B$_2$C (squares) \cite{Yanson}
established by a BTK fit of PC $dV/dI$ curves. Inset: example of a
$dV/dI$ curve (symbols) at 3\,K fitted by BTK theory (thin
curve).}
\end{minipage}\hspace{2pc}%
\begin{minipage}{18pc}
\includegraphics[width=18pc]{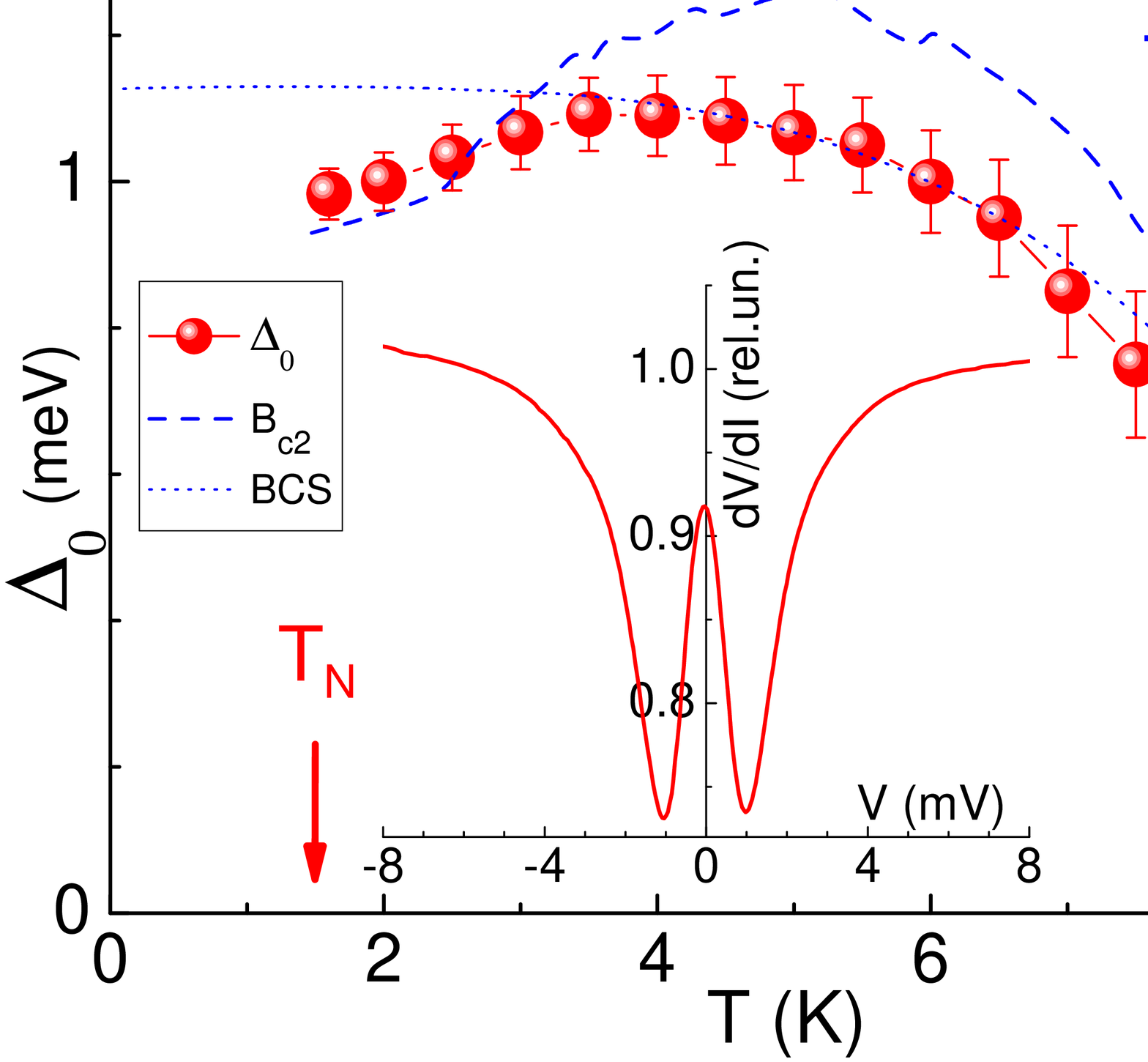}
\vspace{-3cm}\caption{\label{f2}Gap behavior in TmNi$_2$B$_2$C
established by a BTK fit of PC $dV/dI$ curves. The dashed curve
shows qualitatively the B$_{\rm{c2}}$ behavior in TmNi$_2$B$_2$C
along the c-axis. Inset: example of a $dV/dI$ curve at 1.6\,K.}
\end{minipage}
\end{figure}

\section{Results and  discussion}
The SC gap $\Delta _0$ manifests itself on the $dV/dI$
characteristic of a normal metal-superconductor PC by pronounced
minima symmetrically placed at $V \approx \pm \Delta _0$/e if
$T<<$ T$_{\rm c}$ (see inset in Fig.\,1). The measured $dV/dI$ of
the title compounds exhibit one pair of minima as in the case of
single gap superconductors \cite{Naid}, therefore a single gap
approach is usually used to fit experimental data. It is seen from
the inset in Fig.\,1 that the one-gap fit reasonably describes
$dV/dI$ for $R$=Ho. The same is true for $R$=Dy \cite{Yanson} (not
shown). The obtained $\Delta _0(T)$ in Fig.\,1 exhibits a BCS-like
$T$--dependence in both cases, however, $\Delta _0(T)$ vanishes at
$T$*$\approx $5.6\,K for $R$=Ho, a few K below the bulk critical
temperature $T_{\rm c} \approx $ 8.5\,K. It was suggested in
\cite{Naid07} that superconductivity in the commensurate AF phase
in the $R$=Ho compound survives at a special nearly isotropic
Fermi surface sheet, while the gap suppression at $T$* may be
caused by a peculiar magnetic order developing in this compound
above the AF state at $T_{\mbox{\tiny N}}\approx $5.2\,K.

Interesting peculiar behavior of $\Delta_0(T)$ in TmNi$_2$B$_2$C
(Fig.\,2) was found recently \cite{Naid07a}. The SC gap has a
maximum around 4-5\,K and further decreases at lowering
temperature. This is in accord  with the behavior of the upper
critical field along the c-axis. Apparently, AF fluctuations
occurring above the magnetically ordered state at $T_{\mbox{\tiny
N}}$ = 1.5\,K are responsible for the decrease of the SC gap
observed at low temperatures.

As it was shown in \cite{Bobrov08}, the "one-gap" approach to fit
the measured high-quality $dV/dI$ curves\footnote{High quality
means that $dV/dI$ looks like a smooth curve with pronounced AR
minima (see inset in Fig.\,3) while some irregularities like
e.\,g. spike-like structures above the minima (see inset in
Fig.\,1) are absent.} for ErNi$_2$B$_2$C results in a clear
discrepancy between the fit and the data at the minima position
and at zero bias. At the same time, a "two-gap" approach allows a
better fit of $dV/dI$ for ErNi$_2$B$_2$C \cite{Bobrov08}. As it
was mentioned above, the upper critical field $H_{\rm{c2}}(T)$ of
nonmagnetic borocarbides $R$=Y and Lu \cite{Shulga} can be
properly described only by a two-band model. Therefore, the
detection of two SC gaps in magnetic ErNi$_{2}$B$_{2}$C (with
about the same magnitude as in $R$=Lu \cite{Bobrov}) from one hand
testifies for similarities in the electronic band structure of the
mentioned compounds, from the other hand it points to the fact
that superconductivity and magnetism develop in ErNi$_{2}$B$_{2}$C
in different bands.

\vspace{2cm}
\begin{figure}[h]
\begin{minipage}{18pc}
\includegraphics[width=18pc]{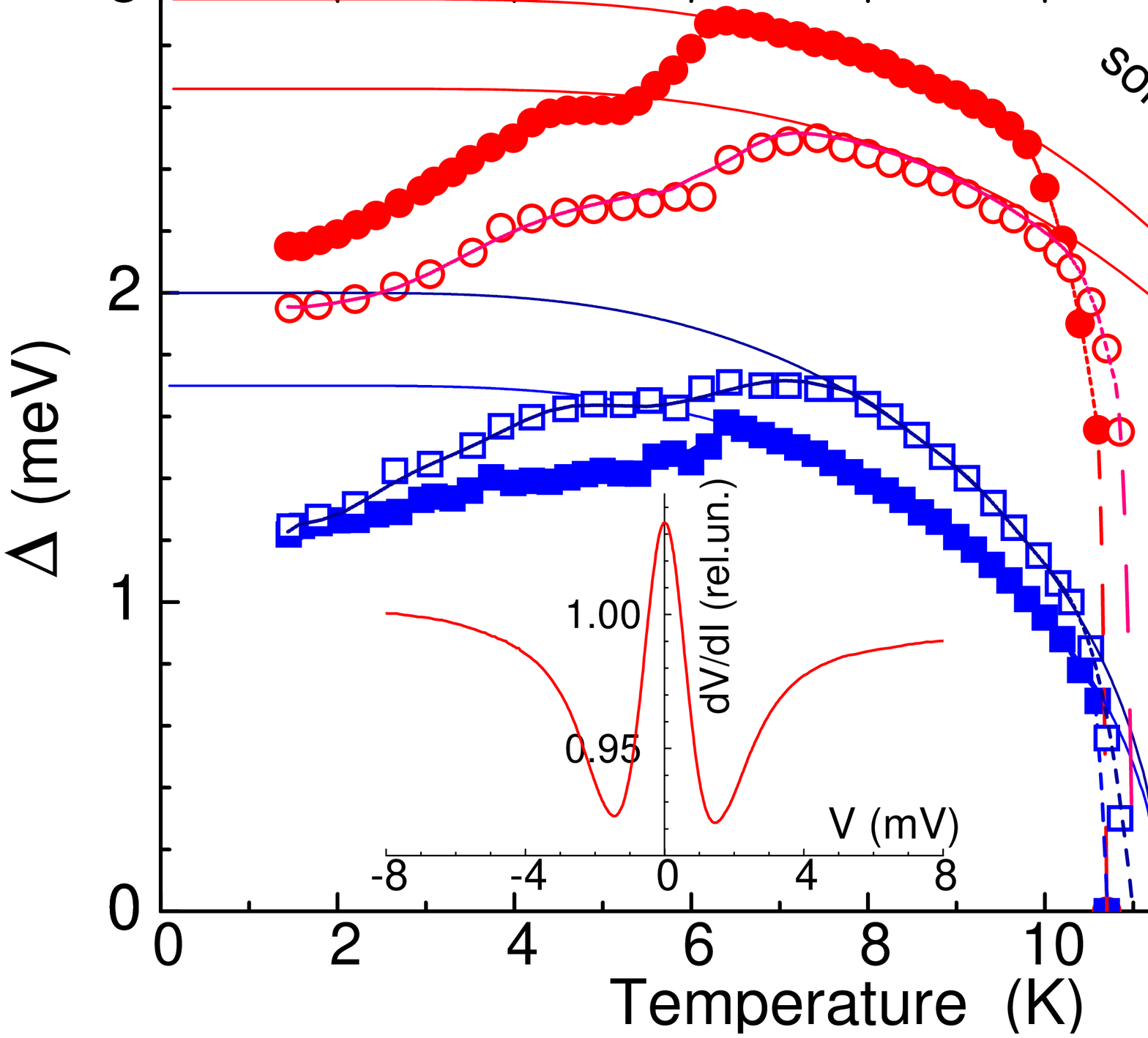}
\vspace{-3cm} \caption{\label{f3}Temperature dependence of the
larger OP (circles) and the smaller OP (squares) for
ErNi$_2$B$_2$C: open symbols -- c-direction, solid -- ab-plane.
Inset: example of a $dV/dI$ curve at 1.46\,K for c-direction
\cite{Bobrov08}.}
\end{minipage}\hspace{1pc}%
\begin{minipage}{18pc}
\includegraphics[width=18pc]{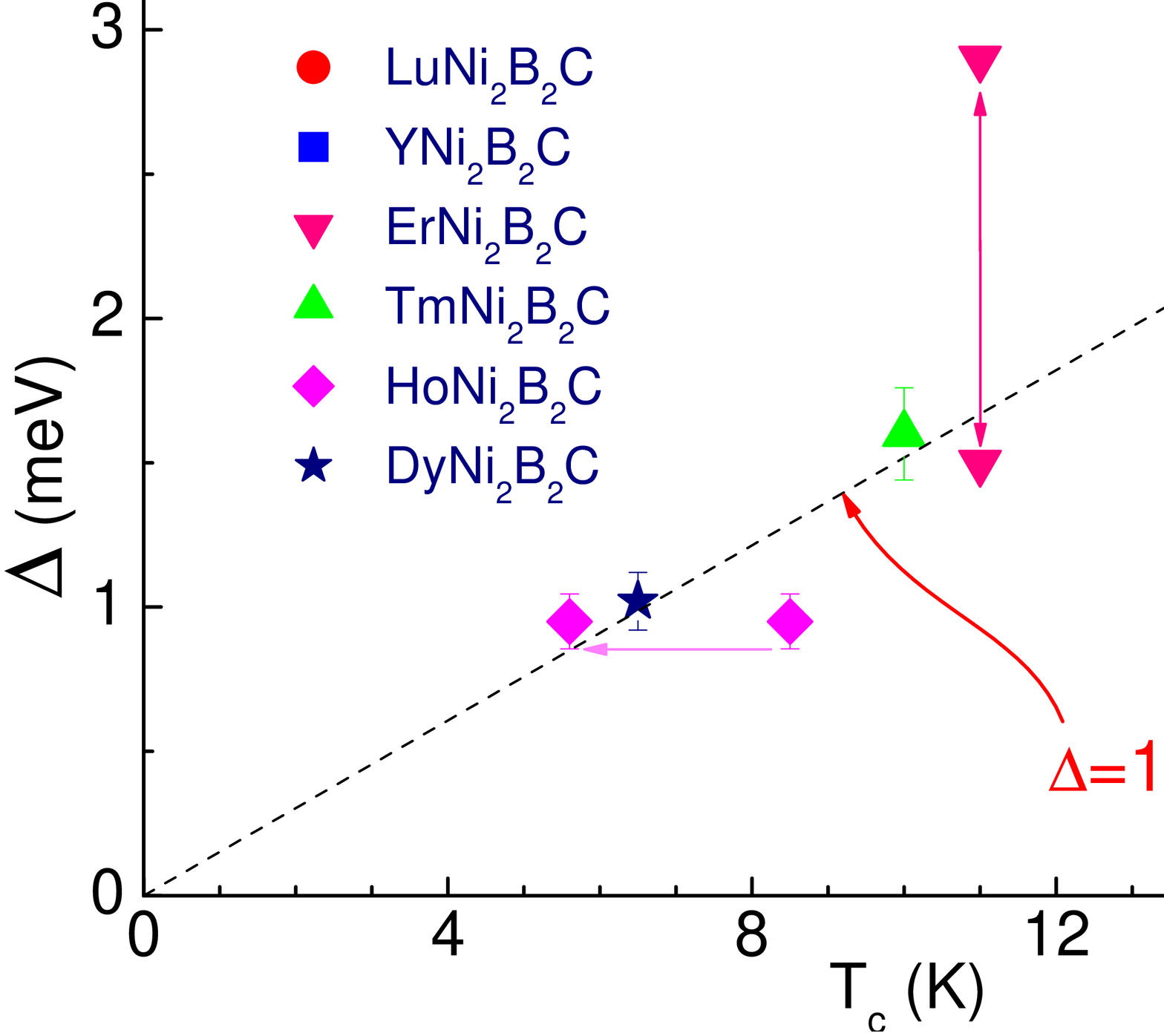}
\vspace{-3cm}\caption{\label{f4}The SC gap or OP (for $R$=Er \&
Lu) established by a PC study vs T$_{\rm c}$ in $R$Ni$_2$B$_2$C
including also the nonmagnetic compounds $R$=Y and Lu.}
\end{minipage}
\end{figure}

From the $T$--dependence of the SC order parameter (OP)
in ErNi$_2$B$_2$C (Fig.\,3) is clearly seen that both OPs start to
decrease by approaching or below T$_{\mbox{\tiny N}}$. A similar
decrease of the SC gap in ErNi$_2$B$_2$C in the AF state was
reported by STM measurements in \cite{Wata} and, recently, by
laser-photoemission spectroscopy \cite{Baba}. Such a gap decrease
in the AF state is also in line with the Machida theory
\cite{Machida} in which spin-density-wave ordering competes with
superconductivity.

On the other hand, $\Delta (T)$ in ErNi$_2$B$_2$C above
$T_{\mbox{\tiny N}}$ is close to the BCS-like behavior, only the
steep vanishing of the larger OP at $T_{\rm c}$ is rather
unexpected. It turned out \cite{Bobrov08} that the contribution of
the larger OP to $dV/dI $ is also temperature dependent decreasing
with increasing temperature and containing only about 10{\%} close
to $T_{\rm c}$. Therefore, it seems that the larger OP disappears
at $T_{\rm c}$ due to a "shrinking" of the corresponding SC part
of the Fermi surface.

It should be noted that the magnetic structure in $R$=Er and Tm
compounds is given by an incommensurate spin density wave. As it
is seen from Fig.\,2, $\Delta _0(T)$ for TmNi$_2$B$_2$C deviates
from the BCS behavior approaching $T_{\mbox{\tiny N}}$. Contrary,
Ho and Dy compounds with commensurate AF order display a BCS-like
gap. Of course, in the case of $R$=Tm measurements below
T$_{\mbox{\tiny N}}\approx $1.5\,K are very desirable to trace
similarity with the Er compound.

Fig.\,4 summarizes the measured SC gap/OP by PCs in the title
compounds. In general, the SC gap/OP values are close to the BCS
value $\Delta _{0}$=1.76 k$_{\rm B}T_{\rm c}$ taking into account
the multiband behavior in the Er--compound and the vanishing of
the SC gap in $R$=Ho at $T^*\approx $5.6\,K. For comparison, also
the SC gaps of the nonmagnetic compounds $R$=Y and Lu are
presented in Fig.\,4. A two-gap state is established for $R$=Lu
\cite{Bobrov} and a strong anisotropy of the gap (probably due to
multiband state) is observed for $R$=Y \cite{Bashlakov}. It is
seen that  the OPs in Er and Lu compounds have close values, but
in the case of Er the larger OP contribution to $dV/dI$ dominates
at low temperature. However, then this contribution decreases (by
a factor 5) with increasing temperature \cite{Bobrov08}, while for
$R$=Lu a similar behavior is observed for the smaller OP
\cite{Bobrov}.

\section{Conclusions}
The SC gap (SC OP) was studied using normal metal-superconductor
PCs for a series of magnetic rare-earth nickel borocarbide
superconductors. For the first time, the existence of two SC OPs
in the magnetic compound ErNi$_{2}$B$_{2}$C has been shown.
Moreover, a distinct decrease of the both OPs is observed as the
temperature is lowered below $T_{\mbox{\tiny N}}$. For the $R$= Ho
and Dy compounds with commensurate AF order, the SC gap has a BCS
like behavior in the AF state, while for $R$=Tm the gap starts to
decrease by approaching a magnetic state with incommensurate AF
order. Note that the $R$=Er compound has an incommensurate AF
order and the OPs also start to decrease slightly above
$T_{\mbox{\tiny N}}$ at lowering temperature. Thus, the
discrepancy in the magnetically ordered state between $R$=Ho, Dy
(commensurate state) and $R$=Er, Tm (incommensurate state) results
in a different SC gap (or OP) behaviors. More extensive
directional PC measurements for $R$=Dy, Ho and Tm are desirable to
check the presence of multi-gap superconductivity in these
compounds as well.

\section{Acknowledgments} The authors thank P. C.
Canfield, K.-H. M\"{u}ller, K. Nenkov, M. Schneider, D. Souptel,
L. V. Tyutrina for the long term collaboration in the field of
rare-earth nickel borocarbides investigations, sample preparation
and experimental assistance. Two of us, Yu. G. N. and O.\ E.\ K.,
thank Prof. L. Schultz and the Alexander von Humboldt Foundation
for support.

\end{document}